\documentclass[%
reprint,
superscriptaddress,
amsmath,amssymb,
aps,
prd,
]{revtex4-1}

\usepackage{graphicx}% Include figure files
\usepackage{dcolumn}% Align table columns on decimal point
\usepackage{bm}% bold math
\begin{document}

\preprint{APS/123-QED}

\title{Calculations of the dominant long-range, spin-independent contributions to the interaction energy between two nonrelativistic Dirac fermions from double-boson exchange of spin-0 and spin-1 bosons with spin-dependent couplings}% Force line breaks with \\
%\thanks{A footnote to the article title}

\author{S. Aldaihan}
\affiliation{Physics Department, Indiana University, Bloomington, Indiana 47408.}%Lines break automatically or can be forced with \\
\author{D. E. Krause}
\affiliation{Physics Department, Wabash College, Crawfordsville, Indiana 47933}
\affiliation{Department of Physics and Astronomy, Purdue University, West Lafayette, Indiana 47907}
\author{J. C. Long }
\affiliation{Physics Department, Indiana University, Bloomington, Indiana 47408.}%Lines break automatically or can be forced with \\
\author{W. M. Snow}%
\affiliation{Physics Department, Indiana University, Bloomington, Indiana 47408.}%Lines break automatically or can be forced with \\

\date{\today}% It is always \today, today,
% but any date may be explicitly specified

\begin{abstract}
Various theories beyond the Standard Model predict new particles with masses in the sub-eV range with very weak couplings to ordinary matter which can possess spin-dependent couplings to electrons and nucleons. Present laboratory constraints on exotic spin-dependent interactions with pseudoscalar and axial couplings for exchange boson masses between meV and eV are very poor compared to constraints on spin-independent interactions in the same mass range arising from spin-0 and spin-1 boson exchange. It is therefore interesting to analyze in a general way how one can use the strong experimental bounds on spin-independent interactions to also constrain spin-dependent interactions by considering higher-order exchange processes. The exchange of a pair of bosons between two fermions with spin-dependent couplings will possess contributions which flip spins twice and thereby generate a polarization-independent interaction energy which can add coherently between two unpolarized objects. In this paper we derive the dominant long-range contributions to the interaction energy between two nonrelativistic spin-1/2 Dirac fermions from double exchange of spin-0 and spin-1 bosons proportional to couplings of the form $g_P^{4}$, $g_S^{2}g_P^{2}$, and $g_V^{2}g_A^{2}$. Our results for $g_P^{4}$ are in agreement with previous calculations that have appeared in the literature. We demonstrate the usefulness of this analysis to constrain spin-dependent couplings by presenting the results of a reanalysis of data from a short-range gravity experiment to derive an improved constraint on $(g^N_{P})^2$, the pseudoscalar coupling for nucleons, in the range between $40$ and $200~\mu$m of about a factor of 5 compared to previous limits. We hope that the expressions derived in this work will be employed by other researchers in the future to evaluate whether or not they can constrain exotic spin-dependent interactions from spin-independent measurements. The spin-independent contribution from 2-boson exchange with axial-vector couplings requires special treatment and will be explored in another paper.

\end{abstract}
%\begin{description}
%\end{description}
%\end{abstract}
\maketitle

%\tableofcontents

\section{\label{sec:level1}Introduction}

The possible existence of new interactions in nature with ranges of mesoscopic scale (millimeters to microns), corresponding to exchange boson masses in the 1 meV to 1 eV range and with very weak couplings to matter has been discussed for some time~\cite{Leitner,Hill} and has recently begun to attract renewed scientific attention. Particles which might mediate such interactions are sometimes referred to generically as WISPs (Weakly-Interacting sub-eV Particles)~\cite{Jae10} in recent theoretical literature. Many theories beyond the Standard Model, including string theories, possess extended symmetries which, when broken at a high energy scale, lead to weakly-coupled light particles with relatively long-range interactions such as axions, arions, familons, and Majorons~\cite{Arvanitaki2010, PDG14}. The well-known Goldstone theorem in quantum field theory guarantees that the spontaneous breaking down of a continuous symmetry at scale $M$ leads to a massless pseudoscalar mode with weak couplings to massive fermions $m$ of order $g=m/M$. The mode can then acquire a light mass (thereby becoming a pseudo-Goldstone boson) of order $m_{boson}=\Lambda^{2}/M$ if there is also an explicit breaking of the symmetry at scale $\Lambda$~\cite{Weinberg72}. New axial-vector bosons such as paraphotons~\cite{Dobrescu2005} and extra Z bosons~\cite{Appelquist2003} appear in certain gauge theories beyond the Standard Model. Several theoretical attempts to explain dark matter and dark energy also produce new weakly-coupled long-range interactions. The fact that the dark energy density of order (1 meV)${^4}$ corresponds to a length scale of $~100$ $\mu$m also encourages searches for new phenomena on this scale~\cite{Ade09}. 

A general classification of interactions between nonrelativistic fermions assuming only rotational invariance~\cite{Dob06} reveals 16 operator structures involving the spins, momenta, interaction range, and various possible couplings of the particles. Of these sixteen interactions, one is spin-independent, six involve the spin of one of the particles, and the remaining nine involve both particle spins. Ten of these 16 possible interactions depend on the relative momenta of the particles. The addition of the spin degree of freedom opens up a large variety of possible new interactions to search for which might have escaped detection to date. Powerful astrophysical constraints on exotic spin-dependent couplings~\cite{Raffelt1995a, Raffelt1995b, Raffelt2012} exist from stellar energy-loss arguments, either alone or in combination with the very stringent laboratory limits on spin-independent interactions from gravitational experiments~\cite{Adelberger:2014}. However, a chameleon mechanism could in principle invalidate some of these astrophysical bounds while having a negligible effect in cooler, less dense lab environments~\cite{Jain2006}, and the astrophysical bounds do not apply to axial-vector interactions~\cite{Dob06}. These potential loopholes in the astrophysical constraints, coupled with the intrinsic value of controlled laboratory experiments and the large range of theoretical ideas which can generate exotic spin-dependent interactions, has led to a growing experimental activity to search for such interactions in laboratory experiments. 

Many experiments search for a monopole-dipole interaction~\cite{Moody84} involving an exchange of spin-zero bosons with scalar and pseudoscalar couplings. This interaction violates $P$ and $T$ symmetry and in the nonrelativistic limit is proportional to $g_{S}g_{P}\vec{\sigma} \cdot \hat{r}$ where $g_{S}$ and $g_{P}$ are the scalar and pseudoscalar couplings, $\vec{\sigma}$ is the spin of one of the particles, and $\vec{r}$ is the separation between the particles. Contrary to some expectations, experimental upper bounds on electric dipole moments, which are also $P$-odd and $T$-odd, do not in general rule out the existence of such bosons with masses in the meV to eV range~\cite{Mantry2014}. Many of the experiments which have been performed to search for such interactions using polarized gases~\cite{You96} and paramagnetic salts~\cite{Chui93, Ni94, Ni99} are sensitive to ranges $\lambda \geq 1$ cm. Constraints on monopole-dipole interactions involving nucleons at smaller range have come from experiments using slow neutrons~\cite{Bae07, Ser09, Ig09, Fed13, Jen14, Afach2015} and polarized helium and xenon gas~\cite{Pok10, Pet10, Fu11, Zhe12, Chu13, Bul13, Tul13, Guigue15}. Many experiments have also sought exotic spin-spin interactions proportional to $g_{P}^{2} \vec{\sigma}_{1} \cdot \vec{\sigma}_{2}$, where $g_{P}$ is the pseudoscalar coupling and $\vec{\sigma}_{1}$ and $\vec{\sigma}_{2}$ are the spins of the two particles. Such a spin-dependent potential with a dipole-dipole form is one of the three velocity-independent spin-spin interactions which can come from 1-boson exchange between two nonrelativistic spin-$1/2$ fermions~\cite{Dob06}. Separated ensembles of polarized atoms~\cite{Wine91, Gle08, Vas09, Hunter2013} have set limits on long-range spin-dependent nucleon interactions, and analysis of high precision spectroscopy in molecular hydrogen~\cite{Ram79, Ledbetter2013} has set limits on atomic-range spin-dependent nucleon interactions. Torsion balance measurements have recently set new stringent limits on both monopole-dipole interactions and dipole-dipole interactions involving polarized electrons with macroscopic ranges~\cite{Rit93, Ham07, Heckel2008, Hoedl11, Heckel2013, Terrano2015}. Comparison of precision QED calculations with atomic physics data~\cite{Karshenboim2011} has set strong limits on exotic spin-dependent electron interactions with ranges at the atomic scale. Ion traps~\cite{Kotler2015} have recently constrained exotic spin-spin interactions between polarized electrons of the form $g_{A}^{2} \vec{\sigma}_{1} \cdot \vec{\sigma}_{2}$ from spin-$1$ boson exchange at micron distance scales. New experimental methods to search for polarized electron couplings using rare earth-based ferrimagnetic test masses~\cite{Leslie2014}, paramagnetic insulators~\cite{Chu2015}, and spin-exchange relaxation-free (SERF) magnetometers~\cite{Chu2016} have been proposed. 

Laboratory constraints on possible new interactions of mesoscopic range which depend on {\it both} the spin {\it and} the relative momentum are less common, since the polarized electrons or nucleons in most experiments employing macroscopic amounts of polarized matter typically possess $\langle\vec{p}\rangle=0$ in the lab frame. Some limits exist for spin-$0$ boson exchange. Kimball {\it et al.}~\cite{Kim10} used measurements and calculations of cross sections for spin exchange collisions between polarized $^{3}$He and Na atoms to constrain possible new spin-dependent interactions between neutrons and protons. Hunter~\cite{Hunter2014} exploited the existence of a small but nonzero polarization of the electrons in the Earth combined with atomic magnetometry to place very stringent constraints on a large number of spin and velocity-dependent interactions involving polarized electrons for macroscopic force ranges. 

Spin and velocity-dependent interactions from spin-$1$ boson exchange can be generated by a light vector boson $X_{\mu}$ coupling to a fermion $\psi$ with an interaction of the form $\mathcal{L}_{I}=\bar{\psi}(g_{V}\gamma^{\mu}+g_{A}\gamma^{\mu}\gamma_{5})\psi X_{\mu}$, where $g_{V}$ and $g_{A}$ are the vector and axial couplings. In the nonrelativistic limit, this interaction gives rise to two interaction potentials of interest depending on both the spin and the relative momentum~\cite{Pie11}: one proportional to $g_{A}^{2}\vec{\sigma}\cdot(\vec{v}\times\hat{r})$ and another proportional to $g_{V}g_{A}\vec{\sigma}\cdot\vec{v}$. As noted above, many theories beyond the Standard Model can give rise to such interactions. For example, spontaneous symmetry breaking in the Standard Model with two or more Higgs doublets with one doublet responsible for generating the up quark masses and the other generating the down quark masses can possess an extra U(1) symmetry generator distinct from those which generate $B$, $L$, and weak hypercharge $Y$. The most general U(1) generator in this case is some linear combination $F=aB + bL +cY + dF_{ax}$ of $B$, $L$, $Y$, and an extra axial U(1) generator $F_{ax}$ acting on quark and lepton fields, with the values of the constants $a,b,c,d$ depending on the details of the theory. The new vector boson associated with this axial generator can give rise to $\mathcal{L}_{I}$ above~\cite{Fayet:1990}. 

Neutrons have recently been used with success to tightly constrain possible weakly coupled spin-dependent interactions of mesoscopic range~\cite{Dubbers11}. A polarized beam of slow neutrons can have a long mean free path in matter and is a good choice for such an experimental search~\cite{Nico05b}. Piegsa and Pignol~\cite{Pie12} recently reported improved constraints on the product of axial vector couplings $g_{A}^{2}$ in this interaction. Polarized slow neutrons which pass near the surface of a plane of unpolarized bulk material in the presence of such an interaction experience a phase shift which can be sought using Ramsey's well-known technique of separated oscillating fields~\cite{Ramsey:1950}. Other experiments have constrained $g_{V}g_{A}^{n}$. Yan and Snow reported constraints on $g_{V}g_{A}^{n}$ using data from a search for parity-odd neutron spin rotation in liquid helium~\cite{Yan13}. Adelberger and Wagner~\cite{Adelberger:2014} combined experimental constraints on $g_{V}^{2}$ from searches for violations of the equivalence principles and $g_{A}^{2}$ from other sources to set much stronger constraints on $g_{V}g_{A}^{n}$ for interactions with ranges beyond 1 cm. Yan~\cite{Yan15} analyzed the dynamics of ensembles of polarized $^{3}$He gas coupled to the Earth to constrain $g_{V}g_{A}^{n}$ for interactions with ranges beyond 1 cm with laboratory measurements.

The strength of nearly all of these constraints is very weak compared to spin-independent interactions. Very stringent constraints exist on spin-independent Yukawa interactions arising from light scalar or vector boson exchange. The present constraints on the dimensionless coupling constants are $g^2_{S,V}$ $ \leq 10^{-40}$ for an exchange boson with a mass between 10 meV and 100 $\mu$eV \cite{Decca}, which corresponds to a length scale between 10 $\mu$m and 1 mm. Experimental constraints on possible new interactions of mesoscopic range which depend on the spin of one or both of the particles are much less stringent than those for spin-independent interactions~\cite{Leslie2014, Antoniadis11}. Several facts contribute to this situation. First of all such experiments require one or both of the particles under investigation to be polarized. Even if one can achieve perfect polarization, only the valence fermions in the ground states of bound electrons and nucleons are accessible. Experimental polarization techniques are often specific to particular atoms or nuclei and vary widely in their efficiency. Macroscopic objects with large nuclear or electron polarization are not easy to arrange without an environment that includes large external magnetic fields. Even if one succeeds to polarize ensembles of particles in low ambient magnetic fields, the magnetic moments of the spin-aligned particles themselves generate magnetic fields which eventually interact with and depolarize other members of the ensemble. Both internal and external magnetic fields can produce large systematic effects in delicate experiments.
Another reason for the differing sensitivities follows from the fact that, for the small momentum transfers accessed in interactions between two nonrelativistic massive Dirac fermions, the amplitude for a helicity flip associated with a spin-dependent interaction at the fermion-boson vertex can be suppressed by a factor $(\mu/m)^n$, where $\mu$ is the mass of the exchanged boson, $m$ is the fermion mass and $n$ = 1, 2, or 3 depending on the type of interaction. This suppression arises at parity-odd vertices such as $i g_{P} \gamma_5$, $g_{V}\boldsymbol{\gamma}$, and $g_{A} {\gamma_0}{\gamma_5}$ where in order for parity to be conserved the boson must be emitted with nonzero angular momentum relative to the initial and final nonrelativistic fermions, thus giving rise to an angular momentum suppression of order $(\mu/m)^n$. The only case of a spin-dependent interaction with no mass suppressions arises in the ``dipole-dipole'' interaction mediated by an axial boson with even-parity coupling $g_{A} \boldsymbol{\gamma}{\gamma_5}$.
This is one of the reasons why, for example the constraint on an electron axial vector coupling $(g^{e}_{A})^2 \sim 10^{-40}$ for $\mu \geq 1$ $\mu$eV \cite{Heckel2013} is orders of magnitude stronger than the constraint on $(g^{N}_{A})^2 \sim 10^{-13}$ for $\mu \sim 100$ $\mu$eV  \cite{Pie12}, where the latter was obtained from a ``monopole-dipole'' interaction arising from parity-odd vertices. \\
The huge difference in the strength of these constraints on spin-dependent and spin-independent interactions motivated us to investigate whether or not limits on spin-dependent couplings can be improved using the constraints from existing spin-independent data. Exchange of two bosons can flip the helicity of the fermions twice and generate a spin-{\it independent} contribution to the interaction energy between two fermions. Although two boson exchange between fermions generates an interaction energy of order $g^{4}$ and direct spin-dependent experiments look for effects from single boson exchange of order $g^{2}$, the strong constraints from spin-independent experiments can still be better than direct experiments in certain situations. Since searches for new spin-independent interactions span a broader range of exchange boson masses than the spin-dependent searches, such an analysis can extend constraints on spin-dependent interactions to new length scales where experimental coverage is either poor or nonexistent. Many experiments to search for spin-independent interactions are probing the smaller distance scales where limits on spin-dependent interactions are poor \cite{kamiya, tanya}.

Similar analyses motivated by the same considerations have been conducted in the past. The functional form for 2-boson exchange with pseudoscalar couplings has been derived before and applied in different contexts~\cite{ferrer,Drell,mostepanenko,grifols} such as tests of the inverse square law of gravity (ISL) and the weak equivalence principle~\cite{2-pseudoscalar} to derive the first direct limits on $g^N_{P}$. The most recent constraints on spin-0 boson exchange with pseudoscalar couplings $g^N_{P}$ to nucleons~\cite{klimchitskaya} span bosons masses between $ 0.01$ $\mu$eV and $1$~eV. 

To the best of our knowledge, no similar analysis has been performed for other spin-dependent couplings and no functional forms for the spin-independent component of the interaction energy arising from other types of 2-boson exchange have been exhibited in the nonrelativistic limit of interest to us. The aim of this paper is to calculate the dominant long-range contribution to the interaction energy between two nonrelativistic spin-1/2 Dirac fermions from double boson exchange of spin-0 and spin-1 bosons with spin-dependent couplings of the form $g_S^{2}g_P^{2}$, and $g_V^{2}g_A^{2}$. The case of two axial vector exchange requires a special treatment and will be explored in another paper. In addition, we use the existing 2-boson calculation for pseudoscalar exchange in a reanalysis of data from a short-range gravity experiment to derive an improved constraint on $(g^N_{P})^2$, the pseudoscalar coupling for nucleons, in the range between $40$ and $200~\mu$m of about a factor of 5 compared to previous limits. This analysis constitutes an existence proof that sensitive experimental searches for spin-independent interactions can also yield the most stringent constraints on spin-dependent interactions at certain distance scales. 

The rest of this paper is organized as follows. In section~\ref{Problem and Method Section} we define the problem and specify the method of calculation. The calculation itself along with the results are outlined in section~\ref{Interaction Energy Section}. In section~\ref{Constraints Section}, we present our derivation of a new limit on nucleon pseudoscalar couplings from analysis of an experiment to probe violations of the inverse square law in short-range gravity. Natural units with $\hbar = c = 1 $ are used throughout the paper. \\

\section{Definition of the Problem and Method} 
\label{Problem and Method Section}

Some groups have undertaken exact calculations of the amplitudes for double boson exchange valid in the relativistic limit~\cite{Guichon}. It is not our purpose here to attempt a complete calculation of this type. We are interested in determining the leading long-range contributions to the spin-independent component of the interaction energy associated with the exchange of two massive spin-0 and spin-1 bosons between two massive spin-1/2 Dirac fermions with various types of spin-dependent couplings. The distance scale regime we are interested in is $r \geqslant \frac{1}{\mu} \gg \frac{1}{m}$, where $r=|\boldsymbol{r}_1-\boldsymbol{r}_2|$ is the separation between the two fermions, $\mu$ is the exchange boson mass, and $m$ is the fermion mass.

Many authors have performed similar calculations for various purposes using different approaches. Iwasaki studied this problem using noncovariant perturbation theory \cite{iwasaki}. Feinberg and Sucher used dispersion methods in covariant perturbation theory~\cite{sucher88} to extract long-range effects from loop corrections. Holstein examined this problem using effective field theory (EFT)~\cite{holstein}. In this paper we shall use a nonrelativistic approach based on ``old fashioned'' perturbation theory (OFPT) using time-ordered diagrams. The reason we are pursuing this approach is that it suffices for the direct identification of spin-independent long-range terms in the nonrelativistic limit that we are interested in. In dispersion methods obtaining long-range effects from loop corrections amounts to calculating $t$-channel discontinuities in Feynman diagrams and performing a Laplace transformation which, although doable in principle, is not necessary for our purposes. A similar procedure could be realized in EFT by recognizing that long-range components are associated with pieces in the scattering amplitude that are non-analytic in momenta transfer~\cite{holstein}. 
\\
\\
We first consider the elastic scattering of two spin-1/2 Dirac fermions of masses $m_1$ and $m_2$. We denote the incoming momenta by $\boldsymbol{p}_1$ and $\boldsymbol{p}_2$ and the outgoing momenta by $\boldsymbol{p}_{1}'$ and $\boldsymbol{p}_{2}'$. The on-shell transition amplitude is given by 
\begin{equation}
T_{fi}(Q) = (2 \pi)^3 \delta( \boldsymbol{p}'_1 + \boldsymbol{p}'_2 - \boldsymbol{p}_1 - \boldsymbol{p}_2) N_f M_{fi}(Q) N_i .
\label{eq:1}
\end{equation} 
where $\boldsymbol{Q}$ is the momentum transfer to the fermion of mass $m$. Here $M_{fi}$ is the Feynman scattering amplitude and $N_f$ and $N_i$ are normalization factors associated with the incoming and outgoing particles in the initial and final states which in the nonrelativistic limit are taken to be unity~\cite{1}.  We define the interaction energy corresponding to the long-range contribution from $M^{(2)}(Q)$ by ~\cite{2}.

\begin{equation}
V^{(2)}(r) = \int \frac{d^{3}Q}{(2\pi)^3} e^{-i\boldsymbol{Q}\cdot \boldsymbol{r}} \: M^{(2)}_{fi}(Q) .
\label{eq:2}
\end{equation}

\section{ Calculation of the Interaction Energy}
\label{Interaction Energy Section}
We start with the Hamiltonian density 
\begin{equation}
H= \overline{\psi}(\boldsymbol{\gamma}\cdot \boldsymbol{p}+m) \psi + H_{\rm int},
\label{eq:4}
\end{equation}
where $\psi$ is the 4-component fermion field. The first term is the free fermion Hamiltonian density and $H_{\rm int}$ is the interaction Hamiltonian density given by 
\begin{equation}
H_{\rm int}=\overline{\psi} [ (g_S + i g_P \gamma_5) \phi + (g_V \gamma^{\mu}+ g_A \gamma^{\mu}\gamma_5) A_{\mu}] \psi,
\label{eq:5}
\end{equation}
where $\phi$ and $A_{\mu}$ are the massive spin-$0$ and spin-$1$ boson fields, respectively. The nonrelativistic limit of the Hamiltonians in Eqs.~\eqref{eq:4} and \eqref{eq:5} are derived by performing a Foldy-Wouthuysen unitary transformation~\cite{cohen} to eliminate all pair production diagrams associated with higher energies which are subdominant in our limit. For our purposes we need only expand the effective Hamiltonian to order $p/m$: 
\begin{subequations}
\begin{eqnarray}
H^{\rm eff}_{S} & = & g_S\psi^+ \psi \phi, \label{Heff S} \\
H^{\rm eff}_{P} & = & \psi^+[-i \frac{g_P}{2m}\boldsymbol{\sigma}\cdot \boldsymbol{k} \phi +\frac{g_P^2}{2m}\phi^2 ]\psi, \label{Heff P} \\
H^{\rm eff}_{V}& = & \psi^+ [g_VA_0-\frac{g_V}{2m}(\pmb{p}+\pmb{p}')\cdot \pmb{A} -i\frac{g_V}{2m}\pmb{\sigma}\cdot \boldsymbol{k}\times \pmb{A} \nonumber \\
&& \mbox{} +\frac{g_V^2}{2m}\pmb{A}^2] \psi, \label{Heff V} \\
H^{\rm eff}_{A}& = & \psi^+ [-g_A \pmb{\sigma}\cdot\pmb{A}+\frac{g_A}{2m}\pmb{\sigma}\cdot(\pmb{p}+\pmb{p}')A_0+\frac{g_A^2}{2m} A_0^2]\psi, \nonumber \\
\label{Heff A}
%\label{eq:6}
\end{eqnarray} 
\end{subequations}
where $\psi$ is now a 2-component fermion field associated with a positive energy spinor. Here $\boldsymbol{p}$ and $\boldsymbol{p}'$ are the incoming and outgoing momenta of the fermion in each vertex, $\boldsymbol{k}$ is the boson momentum, and $\boldsymbol{A}$ and $A_0$ are the space and time  components of the massive spin-1 field, respectively. 
%

%The relevant time-ordered diagrams shown are shown in Fig.~\ref{feynman diagram figure}. We divide them into two sets: 2 diagrams involving two seagull vertices, 6 involving one seagull vertex, and 12 box and crossed-box diagrams. 

In OFPT momentum (but not energy) is conserved at the vertices. The propagator for internal lines $\frac{1}{E_i\: - \: E_n}$, where $E_i$ is the energy of the initial state and $E_n$ is the energy of the intermediate state, is multiplied by a sum over the transverse and longitudinal modes, $\delta_{ij} \: - \: \frac{ k_i k_j}{\mu^2}$ or $-1 + \frac{{\omega}^2}{\mu^2}$, for each massive spin-1 exchange boson present in the diagram. The internal momenta are summed over in the usual way.

We will only derive the spin-independent long-range contributions to the interaction energy arising from the following three cases: exchanges with two pseudoscalar couplings, exchanges with one scalar and one pseudoscalar coupling, and exchanges with one vector coupling and one axial vector coupling. The case of two axial vector exchange requires insertions from higher order corrections in the small momentum expansion of the Hamiltonian and will be explored in detail in another paper. Although exotic spin-$0$ and spin-$1$ boson exchange could appear together in box and cross box diagrams we are not interested in this case for our purposes. 

\begin{figure*}
%\centering
\includegraphics[scale=0.70]{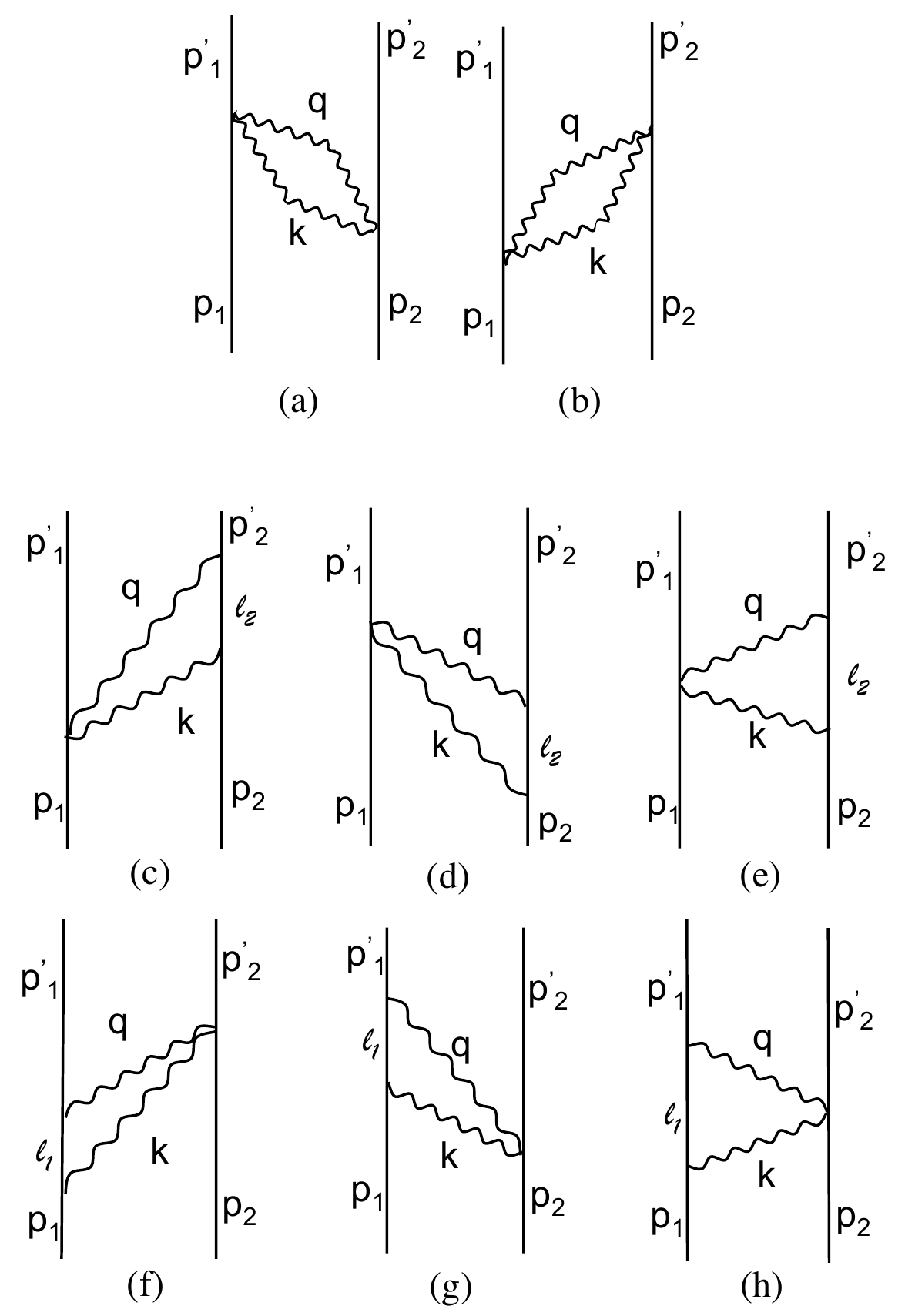}
\caption{\label{feynman diagram figure} The relevant 2-boson exchange time-ordered diagrams. Solid lines represent the fermions while wavy lines represent massive spin-0 or spin-1 bosons.}
\end{figure*}
For exchanges with two pseudoscalar couplings the leading effect comes from the double seagull diagrams (a) and (b) in Fig.~\ref{feynman diagram figure}. Effects arising from diagrams (c)--(h) are suppressed by a factor of $(\mu/m)^3$ as can be inferred from the form of $H^{\rm eff}_{P}$ in Eq.~\eqref{Heff P}. The transition amplitude is 
\begin{eqnarray}
T^{(2)}_{P-P} & =  &- \frac{g^2_{P,1}g^2_{P,2} }{4 m_1 m_2} \: \int \: \frac{d^3k d^3q}{(2\pi)^6} \left[ \frac{1}{ \omega_k \omega_q (\omega_k \:+\: \omega_q) }\right.\nonumber \\
& & \left.{ \delta( \boldsymbol{p}'_1 - \boldsymbol{k} - \boldsymbol{q} - \boldsymbol{p}_1 ) \delta (\boldsymbol{p}'_2 + \boldsymbol{k} + \boldsymbol{q} - \boldsymbol{p}_2) }\right]. 
\label{eq:7}
\end{eqnarray} 
From Eqs.~\eqref{eq:1} and~\eqref{eq:2}, the interaction energy is related to $T^{(2)}_{P-P}$ via 
\begin{equation}
\begin{aligned} 
&{V}^{(2)}_{P-P}(r)= - \frac{g^2_{P,1}g^2_{P,2}}{4 m_1 m_2}\: \int \: \frac{d^{3}Q}{(2\pi)^3} e^{-i\boldsymbol{Q}\cdot \boldsymbol{r}} \\
& \times \int \frac{d^3k d^3q}{(2\pi)^3} \frac{ \delta( \boldsymbol{p}'_1 - \boldsymbol{k} - \boldsymbol{q} - \boldsymbol{p}_1 ) \delta (\boldsymbol{p}'_2 + \boldsymbol{k} + \boldsymbol{q} - \boldsymbol{p}_2) }{ \omega_k \omega_q (\omega_k \:+\: \omega_q) } .
\label{eq:8}
\end{aligned}
\end{equation} 
Now by carrying out the integral over $\boldsymbol{Q}$ first we obtain 
\begin{eqnarray}
{V}^{(2)}_{P-P}(r) &= & - \frac{g^2_{P,1}g^2_{P,2}}{4 m_1 m_2} \: \int \: \frac{d^3k d^3q}{(2\pi)^6} \frac{e^{-i({\boldsymbol{k} + \boldsymbol{q})\cdot \boldsymbol{r}}}}{ \omega_k \omega_q (\omega_k \:+\: \omega_q) } \nonumber \\
&= & -\frac{g^2_{P,1}g^2_{P,2}}{4m_1 m_2}\frac{\mu K_1(2\mu r)}{8 \pi^3 r^2},
\label{eq:9}
\end{eqnarray}
where $K_1(x)$ is the modified Bessel function of the second kind. This agrees with the result previously derived by Drell and Huang \cite{Drell} and Ferrer and Nowakowski~\cite{ferrer}. This result, however, is not correct for the exchange of two pseudoscalar bosons which have a derivative coupling of the form $ \frac{g_P}{m} \overline{\psi} \gamma_{\mu} \gamma_5 \psi \partial^{\mu} \phi $. Derivative and non-derivative pseudoscalar couplings give the same interaction energy in first order perturbation theory but not on second order. The long-range behavior arising from two massless boson exchange with pseudoscalar derivative couplings to matter have been calculated in the limit as the exchange boson mass goes to zero and shown to be highly suppressed relative to the analogous case with non-derivative pseudoscalar couplings\cite{grifols}. This is also expected to follow for non-massless bosons, but we have not calculated this case in this paper. Since the case of pseudoscalar boson exchange is especially interesting from a physics point of view we plan to calculate this case and present the results in a later paper.

For interactions with one scalar coupling and one pseudoscalar coupling, the leading spin-independent contribution arises from diagrams (c)--(h) of Fig.~\ref{feynman diagram figure} with two orders of $g_{S} \phi$ and one order of $(g^2_{P}/2m) \phi^2$. The transition amplitude is 

\begin{widetext}
\begin{eqnarray}
T^{(2)}_{S-P} &= &\frac{g^2_{S,1}g^2_{P,2}}{2 m_2} \int \frac{d^3k d^3q d^3l_1}{(2 \pi)^6} \Bigg\{\frac{1}{ 4 \omega_k \omega_q} \Bigg[ \frac{\delta(\boldsymbol{p}'_2 + \boldsymbol{k} + \boldsymbol{q} - \boldsymbol{p}_2) \delta(\boldsymbol{p}'_1 -\boldsymbol{q} - \boldsymbol{l}_1) \delta(\boldsymbol{l}_1 -\boldsymbol{k} - \boldsymbol{p}_1)}{( \omega_q+X_1)(\omega_k \:+\: \omega_q)}+ \nonumber \\
&& 
\frac{\delta(\boldsymbol{p}'_2 - \boldsymbol{k} - \boldsymbol{q} - \boldsymbol{p}_2) \delta(\boldsymbol{p}'_1 +\boldsymbol{q} - \boldsymbol{l}_1) \delta(\boldsymbol{l}_1 +\boldsymbol{k} - \boldsymbol{p}_1)}{ (\omega_k+X_1)(\omega_k \:+\: \omega_q)} + \frac{\delta(\boldsymbol{p}'_2 + \boldsymbol{q} - \boldsymbol{k} - \boldsymbol{p}_2) \delta(\boldsymbol{p}'_1 -\boldsymbol{q} - \boldsymbol{l}_1) \delta(\boldsymbol{l}_1 +\boldsymbol{k} - \boldsymbol{p}_1)}{ (\omega_k+X_1)( \omega_q+X_1)} \Bigg] \nonumber \\
&& \mbox{} + 1 \leftrightarrow 2 , \boldsymbol{k} \leftrightarrow \boldsymbol{-k}, \boldsymbol{q} \leftrightarrow \boldsymbol{-q} \}.
\label{eq:18}
\end{eqnarray}
\end{widetext}
Expanding in the limit $X \ll \omega_k$ and taking advantage of symmetry under $\boldsymbol{k}$ and $\boldsymbol{q}$ gives
\begin{eqnarray}
T^{(2)}_{S-P} &=& {\frac{g^2_{S,1}g^2_{P,2}}{2 m_2} }  \int \frac{d^3k d^3qd^3l_1}{(2\pi)^6}  \Bigg\{\frac{1}{ 4 \omega_k \omega_q} \nonumber \\
&& \delta(\boldsymbol{p}'_2 + \boldsymbol{k} + \boldsymbol{q} - \boldsymbol{p}_2) \delta(\boldsymbol{p}'_1 -\boldsymbol{q} - \boldsymbol{l}_1) \delta(\boldsymbol{l}_1 -\boldsymbol{k} - \boldsymbol{p}_1) \nonumber \\
&& \Bigg[ \frac{1}{\omega_q(\omega_k \:+\: \omega_q)}+\frac{1}{ \omega_k(\omega_k \:+\: \omega_q)} + \frac{1}{ \omega_k \omega_q} \Bigg] \nonumber \\
&& \mbox{}+ 1 \leftrightarrow 2 , \boldsymbol{k} \leftrightarrow \boldsymbol{-k}, \boldsymbol{q} \leftrightarrow \boldsymbol{-q}\}. 
\label{eq:19}
\end{eqnarray}
The interaction energy is then given by 
\begin{equation} 
{V}^{(2)}_{S-P}(r)= \Bigg({\frac{g^2_{S,1}g^2_{P,2}}{2 m_2} + \frac{g^2_{S,2}g^2_{P,1}}{2 m_1} }\Bigg) \frac{e^{-2 \mu r}}{32 \pi^2 r^2}.
\label{eq:20}
\end{equation} 
The leading spin-independent contribution for the case of one vector coupling and one axial vector coupling also follows from diagrams (c)--(h) of Fig.~\ref{feynman diagram figure}. Two different processes give rise to this interaction at this order: one from two factors of $- g_{A} {\boldsymbol{\sigma}} \cdot {\boldsymbol{A}} $ with one factor of $(g^2_{V}/2m)$ ${\boldsymbol{A}^2}$ and the other from two factors of $g_{V} A_0$ with one factor of $(g^2_{A}/2m) A^2_0$. In the limit $X \ll \omega_k$, the vector-axial interaction energy is given by
\begin{equation}
\begin{aligned}
&{V}^{(2)}_{V-A}(r)=  \: \int \: \frac{d^3k d^3q}{(2 \pi)^6} \Bigg[ \Bigg( {\frac{g^2_{V,1}g^2_{A,2}}{2 m_1} } +{\frac{g^2_{V,2}g^2_{A,1}}{2 m_2} } \Bigg) \frac{\boldsymbol{k}^2 \boldsymbol{q}^2}{\mu^4}+ \\
&\Bigg( {\frac{g^2_{V,2}g^2_{A,1}}{2 m_1} } + {\frac{g^2_{V,1}g^2_{A,2}}{2 m_2} }\Bigg) \Bigg({3-\frac{\boldsymbol{q}^2}{\mu^2}-\frac{\boldsymbol{k}^2}{\mu^2}+\frac{(\boldsymbol{k}\cdot \boldsymbol{q})^2}{\mu^4} } \Bigg) \Bigg]\\ 
& \frac{e^{-i({\boldsymbol{k} + \boldsymbol{q})\cdot \boldsymbol{r}}} }{ 2 \omega_k \omega_q}\Bigg[ \frac{1}{\omega_q(\omega_k \:+\: \omega_q)}+\frac{1}{ \omega_k(\omega_k \:+\: \omega_q)} + \frac{1}{ \omega_k \omega_q} \Bigg].
\label{eq:21}
\end{aligned}
\end{equation}

Integration over $\boldsymbol{k}$ and $\boldsymbol{q}$ gives
\begin{eqnarray}
{V}^{(2)}_{V-A}(r) & = &  \Bigg[ {\frac{g^2_{V,1}g^2_{A,2}}{2 m_1} }+{\frac{g^2_{V,2}g^2_{A,1}}{2 m_2} } \nonumber \\
&& \mbox{}+ 2 \Bigg( {\frac{g^2_{V,2}g^2_{A,1}}{2 m_1} }+ {\frac{g^2_{V,1}g^2_{A,2}}{2 m_2} }\Bigg) \nonumber \\
 &&\Bigg( 3+ \frac{2}{\mu r}+ \frac{5}{(\mu r )^2} + \frac{6}{(\mu r )^3} + \frac{3}{(\mu r )^4} \Bigg)\Bigg] \nonumber \\ 
 && \times \frac{e^{-2 \mu r}}{ 16 \pi^2 r^2},
\label{eq:22}
\end{eqnarray}
which is the same as Eq.~\eqref{eq:20} except for extra terms due to the sum over polarization states. These terms possess singularities in the $\mu \rightarrow 0$ limit due to the inclusion of the longitudinal component of the massive spin-1 field in the absence of a conserved current~\cite{Karshenboim2011, malta, Fischback, Fayet}. As we never let $\mu \to 0$ by assumption this infrared singularity is not realized in our case. The range of validity of Eq.~(\ref{eq:22}) is $r \gg 1/\mu \gg 1/m_1, 1/m_2$ with $\mu$ finite, in which case it simplifies to
\begin{eqnarray}
{V}^{(2)}_{V-A}(r) & \simeq & \left[ {\frac{g^2_{V,1}g^2_{A,2}}{2 m_1} }+{\frac{g^2_{V,2}g^2_{A,1}}{2 m_2} } \right. \nonumber \\
&& \mbox{} + \left. 6 \left( {\frac{g^2_{V,2}g^2_{A,1}}{2 m_1} }+ {\frac{g^2_{V,1}g^2_{A,2}}{2 m_2} }\right)\right] \frac{e^{-2 \mu r}}{ 16 \pi^2 r^2}.
\phantom{spa}
\label{eq:23}
\end{eqnarray}
\section{Constraints from the Indiana Short-Range Gravity Experiment} 
\label{Constraints Section}

To illustrate the potential power of these results, we have used existing data
from a previous short-range gravity experiment to constrain the couplings in
the interaction energies in Eqs.~\eqref{eq:9}, \eqref{eq:20}, and \eqref{eq:23}. 
This experiment is optimized for sensitivity to macroscopic, spin-independent
forces beyond gravity at short range, which in turn could
arise from exotic elementary particles or even extra spacetime
dimensions. It is described in detail elsewhere~\cite{Long03,Yan14};
here we concentrate on the essential features. 

The experiment is illustrated in Fig.~\ref{feynman diagram figure} of Ref.~\cite{Long15}. The
test masses consist of 250~$\mu$m thick planar tungsten 
oscillators, separated by a gap of 100~$\mu$m, with a stiff conducting
shield in between them to suppress 
electrostatic and acoustic backgrounds. Planar geometry
concentrates as much mass as possible at
the scale of interest, and is nominally null with
respect to $1/r^{2}$ forces. This is effective in suppressing the
Newtonian background relative to exotic short-range effects, and is
well-suited for testing interactions of the form $e^{-\mu r}$, and
$K_{1}(\mu r)$. The force-sensitive
``detector'' mass is driven by the force-generating ``source'' mass
at a resonance near 1~kHz, placing a heavy burden on
vibration isolation. The 1~kHz operation is chosen since at this
frequency it is possible to construct a simple vibration
isolation system. This design has proven effective for suppressing all
background forces to the level of the thermal noise due to dissipation
in the detector mass~\cite{Yan14}. After a run in 2002, the experiment set the
strongest limits on forces beyond gravity between 10 and
100~$\mu$m~\cite{Long03}. The experiment has since been optimized to
explore gaps below 50~$\mu$m, and new force data were acquired in
2012. These data have been used to set limits on Lorentz invariance
violation in gravity~\cite{Long15, Shao16}.
\begin{figure}[t]
\centering
\includegraphics[width=3.5in]{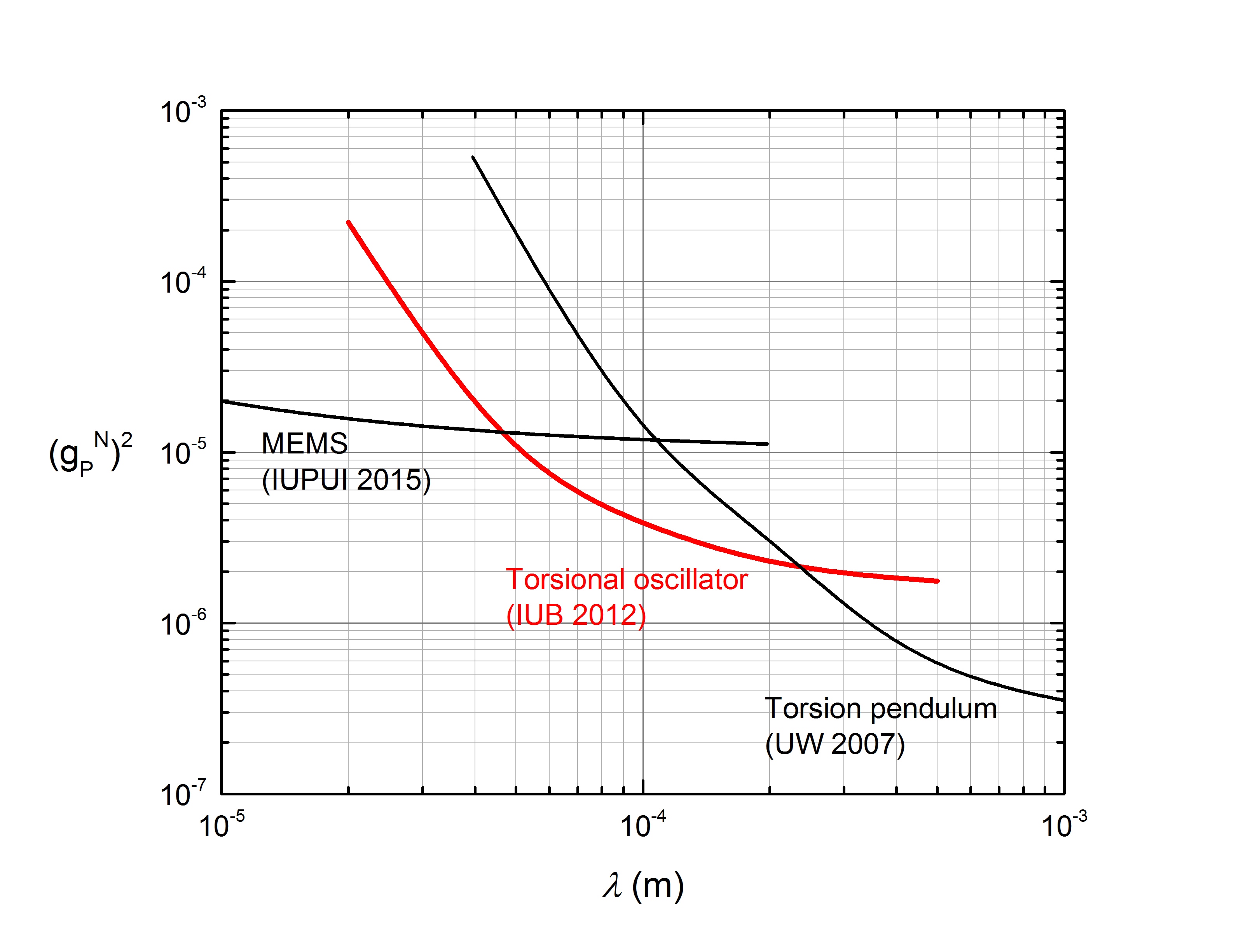}
\caption{\label{fig:limits} Limits on the pseudoscalar coupling for nucleons. Red dashed curve is from this work. Black solid and black dashed curves follow from Ref.\ \cite{2-pseudoscalar} and  \cite{klimchitskaya}, respectively. }
\end{figure}
Analysis of the 2012 data for evidence of double boson exchange
follows that in Ref.~\cite{Long03} for Yukawa-type mass-coupled
forces. Eqs.~\eqref{eq:9}, \eqref{eq:20}, and \eqref{eq:23} are converted to forces and
integrated numerically by Monte Carlo techniques over the 2012
experimental geometry, using the parameters in Refs.~\cite{Long03}
and~\cite{Long15} and their errors. Systematic errors from the
dimensions and positions of the test masses are determined at this stage, by
computing a population of force values generated from a spread of
geometries based on the metrology errors. Gaussian likelihood
functions for the experiment are constructed using the difference
between the measured force and the numerical expressions for the
double exchange forces as the means.

Limits on the double boson exchange interactions are determined by
integration of the likelihood functions over the spin-dependent
couplings (which are free parameters in the likelihood functions), for
several values of the range $\lambda = 1/\mu$. Results for the
2$\sigma$ limits on the coupling $(g_{P}^{N})^{2}$ in Eq.~\eqref{eq:9} are shown
in Fig.~\ref{fig:limits}. The constraints are more sensitive than previously
published limits \cite{2-pseudoscalar, klimchitskaya} by about a factor of 5 in the range near
100~$\mu$m. Analysis of Eqs.~ \eqref{eq:20}, and \eqref{eq:23} is still in progress with the understanding that \eqref{eq:23} is only applicable for $\mu r \gg 1$.  

\section{Conclusion}
We have derived the leading-order spin-independent contribution to the interaction energy arising from the exchange of two light massive bosons between two spin-1/2 Dirac fermions in the nonrelativistic limit. Our expressions agree with previous calculations in the literature where they exist. The functional forms derived in this paper open up an opportunity to constrain, using existing spin-independent data, spin-dependent couplings over new length scales that are outside the sensitivity of current spin-dependent experiments. We also used our expressions to reanalyze data from a short-range gravity experiment. From this analysis we derive a new limit on pseudoscalar couplings for nucleons which is more sensitive than direct constraints from other existing spin-dependent experiments. These limits can be further improved by reconfiguring existing experiments to make them more sensitive to the 2-BEP functional forms.

\begin{acknowledgments}
The work of S. A., J. C. L., and W. M. S. was supported
by the U.S. National Science Foundation Grants No. PHY-
1306942 and No. PHY-1614545, and by the Indiana
University Center for Spacetime Symmetries. S. A.
acknowledges support from a King Abdullah Fellowship.
We also thank E. Fischbach for useful discussions.

\end{acknowledgments}

\bibliography{2bep-paper-PRD-Draft2}% Produces the bibliography via BibTeX.

\begin{thebibliography}{xx}

\bibitem{Leitner} J.\ Leitner and S.\ Okubo, Phys.\ Rev.\ {\bf 136}, B1542 (1964).
\bibitem{Hill} C.T.\ Hill and G. G.\ Ross, Nucl.\ Phys.\ B {\bf 311}, 253 (1988).
\bibitem{Jae10} J. Jaeckel and A. Ringwald, Annu. Rev. Nucl. Part. Sci. {\bf 60}, 405 (2010).
\bibitem{Arvanitaki2010} A.\ Arvanitaki, S.\ Dimopoulos, S.\ Dubovsky, N.\ Kapoler, and J.\ March-Russell, Phys.\ Rev.\ D {\bf 81}, 123530 (2010). 
\bibitem{PDG14} K. A.\ Olive {\it et al.} (Particle Data Group), Chin.\ Phys.\ C {\bf 38}, 090001 (2014).
\bibitem{Weinberg72} S.\ Weinberg, Phys.\ Rev.\ Lett.\ {\bf 29}, 1698 (1972).
\bibitem{Dobrescu2005} B. A.\ Dobrescu, Phys.\ Rev.\ Lett.\ {\bf 94}, 151802 (2005).
\bibitem{Appelquist2003} T.\ Appelquist, B. A.\ Dobrescu, and A.R.\ Hopper, Phys.\ Rev.\ D {\bf 68}, 035012 (2003).
\bibitem{Ade09} E. G.\ Adelberger {\it et al.}, Prog.\ Part.\ Nucl.\ Phys.\ {\bf 62}, 102 (2009).
\bibitem{Dob06} B. Dobrescu and I. Mocioiu, J. High Energy Phys. {\bf 11}, 005 (2006).
\bibitem{Raffelt1995a} G. G.\ Raffelt and A.\ Weiss, Phys.\ Rev.\ D {\bf 51}, 1495 (1995).
\bibitem{Raffelt1995b} G. G.\ Raffelt, {\it Stars as Laboratories for Fundamental
Physics}, University of Chicago Press (1995).
\bibitem{Raffelt2012} G. G.\ Raffelt, Phys.\ Rev.\ D {\bf 86}, 015001 (2012).
\bibitem{Adelberger:2014} E. G.\ Adelberger and T. A.\ Wagner, Phys.\ Rev.\ D {\bf 88}, 031101(R) (2014).
\bibitem{Jain2006} P.\ Jain and S.\ Mandal, Int.\ J.\ Mod.\ Phys.\ D {\bf 15}, 2095 (2006).
\bibitem{Moody84} J. E.\ Moody and F.\ Wilczek, Phys.\ Rev.\ D {\bf 30}, 130 
(1984).
\bibitem{Mantry2014} S.\ Mantry, M.\ Pitschmann, M. J.\ Ramsey-Musolf,
Phys.\ Rev.\ D{\bf 90}, 054016 (2014). 
\bibitem{You96} A. N.\ Youdin, D.\ Krause, K.\ Jagannathan, L. R.\ Hunter, S. K.\ Lamoreaux, Phys.\ Rev.\ Lett.\ {\bf 77}, 2170 (1996).
\bibitem{Chui93} T. C. P.\ Chui and W.-T.\ Ni, Phys.\ Rev.\ Lett. {\bf 71}, 3247 (1993).
\bibitem{Ni94} W.-T.\ Ni, T. C. P.\ Chui, S.-S.\ Pan, and B.-Y.\ Cheng, Physica B: Condensed Matter {\bf 194-196}, 153 (1994). 
\bibitem{Ni99} W.-T.\ Ni, S.-S.\ Pan, H.-C.\ Yeh, L.-S.\ Hou, and J.\ Wan, Phys.\ Rev.\ Lett.\ {\bf 82}, 2439 (1999).
\bibitem{Bae07} S.\ Baessler, V. V.\ Nesvizhevsky, K. V.\ Protasov, and A. Y.\ Voronin, Phys.\ Rev.\ D {\bf 75}, 075006 (2007).
\bibitem{Ser09} A.\ Serebrov, Physics Letters B {\bf 680}, 423 (2009).
\bibitem{Ig09} V. K.\ Ignatovich and Y. N.\ Pokotilovski, Eur.\ Phys.\ J.\ C {\bf 64}, 19 (2009).
\bibitem{Fed13} V. V.\ Fedorov, I. A.\ Kuznetsov, and V. V.\ Voronin, Nucl.\ Inst.\ Meth.\ {\bf B309},  237 (2013). 
\bibitem{Jen14} T.\ Jenke {\it et al.}, Phys.\ Rev.\ Lett. {\bf 112}, 151105 (2014).
\bibitem{Afach2015} S.\ Afach {\it et al.}, Phys.\ Lett.\ B {\bf 745}, 58 (2015). 
\bibitem{Pok10} Y. N.\ Pokotilovski, Phys.\ Lett.\ B {\bf 686}, 114 (2010).
\bibitem{Pet10} A. K.\ Petukhov, G.\ Pignol, D.\ Jullien, and K. H.\ Andersen, Phys.\ Rev.\ Lett.\ {\bf 105}, 170401 (2010).
\bibitem{Fu11} C. B.\ Fu, T. R.\ Gentile, and W. M.\ Snow, Phys.\ Rev.\ D{\bf 83}, 031504(R) (2011).
\bibitem{Zhe12} W.\ Zheng, H.\ Gao, B.\ Lalremruata, Y.\ Zhang, G.\ Laskaris, W.M.\ Snow, and C.B.\ Fu, Phys.\ Rev.\ D {\bf 85}, 031505(R) (2012).
\bibitem{Chu13} P. H.\ Chu {\it et al.}, Phys.\ Rev.\ D {\bf 87}, 011105(R) (2013).
\bibitem{Bul13} M.\ Bulatowicz {\it et al.}, Phys.\ Rev.\ Lett.\ {\bf 111}, 100800 (2013).
\bibitem{Tul13} K.\ Tullney {\it et al.}, Phys.\ Rev.\ Lett.\ {\bf 111}, 100801 (2013).
\bibitem{Guigue15} M.\ Guigue, D.\ Jullien, A. K.\ Petukhov, and G\. Pignol, Phys.\ Rev.\ D {\bf 92}, 114001 (2015). 

\bibitem{Wine91} D. J.\ Wineland, J. J.\ Bollinger, D. J.\ Heinzen, W. M.\ Itano, and M. G.\ Raizen, Phys.\ Rev.\ Lett.\ {\bf 67}, 1735 (1991).
\bibitem{Gle08} A. G.\ Glenday, C. E.\ Cramer, D. F.\ Phillips, and R. L.\ Walsworth, Phys.\ Rev.\ Lett.\ {\bf 101}, 261801 (2008).
\bibitem{Vas09} G.\ Vasilakis, J. M.\ Brown, T. W.\ Kornack, and M. V.\ Romalis, Phys.\ Rev.\ Lett.\ {\bf 103}, 261801 (2009).
\bibitem{Hunter2013} L. R.\ Hunter, J. E.\ Gordon, S. K.\ Peck, D.\ Ang, and J.- F.\ Lin, Science {\bf 339}, 5 (2013).

\bibitem{Ram79} N. F.\ Ramsey, Physica A {\bf 96}, 285 (1979).
\bibitem{Ledbetter2013} M. P.\ Ledbetter, M. V.\ Romalis, and D. F. Jackson Kimball, Phys.\ Rev.\ Lett.\ {\bf 110}, 040402 (2013). 

\bibitem{Ham07} G. D.\ Hammond, C. C.\ Speake, C.\ Trenkel, and A. P.\ Paton, Phys.\ Rev.\ Lett.\ {\bf 98}, 081101 (2007).
\bibitem{Rit93} R. C.\ Ritter, L. I.\ Winkler, and G. T.\ Gillies, Phys.\ Rev.\ Lett.\ {\bf 70}, 701 (1993).
\bibitem{Heckel2008} B. R.\ Heckel {\it et al.}, Phys.\ Rev.\ D {\bf 78}, 092006 (2008).
\bibitem{Hoedl11} S. A.\ Hoedl, F.\ Fleischer, E.G.\ Adelberger, and B. R.\ Heckel, Phys.\ Rev.\ Lett.\ {\bf 106}, 041801 (2011).
\bibitem{Heckel2013} B. R.\ Heckel, W. A.\ Terrano and E. G.\ Adelberger, Phys.\
Rev.\ Lett.\ {\bf 111}, 151802 (2013).
\bibitem{Terrano2015} W. A.\ Terrano, E. G.\ Adelberger, J. G.\ Lee, and B. R.\ Heckel, Phys.\ Rev.\ Lett.\ {\bf 115}, 201801 (2015).

\bibitem{Karshenboim2011} S. G.\ Karshenboim, Phys.\ Rev.\ A {\bf 83}, 062119 (2011).
\bibitem{Kotler2015} S.\ Kotler, R.\ Ozeri, and Derek F.\ Jackson Kimball, Phys.\ Rev.\ Lett.\ {\bf 115}, 081801 (2015).

\bibitem{Leslie2014}
T. M.\ Leslie, E.\ Weisman, R.\ Khatiwada, and J. C.\ Long, Phys.\ Rev.\ D {\bf 89}, 114022 (2014).

\bibitem{Chu2015} P.-H.\ Chu, E.\ Weisman, C.-Y.\ Liu, and J. C.\ Long, Phys.\ Rev.\ D{\bf 91} 102006 (2015). 
\bibitem{Chu2016} P. -H.\ Chu, Y. J.\ Kim, and I.\ Savukov, arXiv: 1606.01152 (2016). 

\bibitem{Kim10} D. F.\ Jackson Kimball, A.\ Boyd, and D.\ Budker, Phys.\ Rev.\ A {\bf 82}, 062714 (2010).

\bibitem{Hunter2014} L. R.\ Hunter and D. G.\ Ang, Phys.\ Rev.\ Lett.\ {\bf 112}, 091803 (2014).

\bibitem{Pie11} F. M.\ Piegsa and G.\ Pignol, Journal of Physics {\bf 340}, 012043 (2012).

\bibitem{Fayet:1990} P.\ Fayet, Nucl.\ Phys.\ B {\bf 347}, 743 (1990).

\bibitem{Dubbers11} D.\ Dubbers and M.\ Schmidt, Rev.\ Mod.\ Phys.\ {\bf 83}, 1111 (2011).
\bibitem{Nico05b} J. S.\ Nico and W. M.\ Snow, Ann.\ Rev.\ Nucl.\ Part.\ Sci.\ {\bf 55}, 27 (2005).
\bibitem{Pie12} F. M.\ Piegsa and G.\ Pignol, Phys.\ Rev.\ Lett.\ {\bf 108}, 181801 (2012).
\bibitem{Ramsey:1950} N. F.\ Ramsey, Phys.\ Rev.\ {\bf 78}, 695 (1950).
\bibitem{Yan13} H.\ Yan and W. M.\ Snow, Phys.\ Rev.\ Lett.\ {\bf 110}, 082003 (2013).
\bibitem{Yan15} H. Y.\ Yan, G. A.\ Sun, S. M.\ Peng, Y.\ Zhang, C.\ Fu, H.\ Guo, and B. Q.\ Liu, Phys.\ Rev.\ Lett.\ {\bf 115}, 182001 (2015). 

\bibitem{Decca}
Y.-J.\ Chen, W.K.\ Tham, D.E.\ Krause, D.\ Lopez, E.\ Fischbach, and R.S.\ Decca,
Phys.\ Rev.\ Lett.\ {\bf 116}, 221102 (2016).

\bibitem{Antoniadis11}
I.\ Antoniadis {\it et al.}, C.\ R.\ Physique {\bf 12}, 755 (2011).

\bibitem{kamiya}
Y. Kamiya, K. Itagaki, M. Tani, G. N. Kim, and S. Komamiya  Phys. Rev. Lett. {\bf 114}, 161101 (2015).


\bibitem{tanya}
Tanya Zelevinsky, private communication (2016).  


\bibitem{ferrer}
F.\ Ferrer and M.\ Nowakowski,
Phys.\ Rev.\ D {\bf 59}, 075009 (1999).

\bibitem{Drell}
S. D.\ Drell and K.\ Huang,
Phys.\ Rev.\ {\bf 91}, 6 (1953).

\bibitem{mostepanenko}
V. M.\ Mostepanenko and Sokolov,
Sov.\ J.\ Nucl.\ Phys.\ {\bf 46}, 685 (1988). 

\bibitem{grifols}
F.\ Ferrer and J.A.\ Grifols,
Phys.\ Rev.\ D {\bf 58}, 096006 (1998).

\bibitem{2-pseudoscalar}
E.\ Fischbach and D.E.\ Krause, 
Phys.\ Rev.\ Lett.\ {\bf 82}, 4753 (1999); 
E.\ Fischbach and D.E.\ Krause, 
Phys.\ Rev.\ Lett.\ {\bf 83}, 3593 (1999); 
E.G.\ Adelberger et.\ al.\ , 
Phys.\ Rev.\ D {\bf 68}, 062002 (2003);
E.G.\ Adelberger, B.R.\ Heckel, S.\ Hoedl, C.D.\ Hoyle, D.J.\ Kapner and A.\ Upadhye,
Phys. Rev. Lett. {\bf 98}, 131104 (2007).

\bibitem{klimchitskaya}
L.\ Klimchitskaya, and V.M.\ Mostepanenko,
Eur.\ Phys.\ J.\ C, {\bf 75}, N4, 164 (2015).

\bibitem{Guichon} 
P. A. M.\ Guichon and M.\ Vanderhaeghen,
Phys. Rev. Lett. 91, 142303 (2003);
P. G.\ Blunden, W.\ Melnitchouk and J. A.\ Tjon, 
Phys.\ Rev.\ Lett.\ {\bf 91}, 142304 (2003);
A. V.\ Afanasev, S. J.\ Brodsky, C. E.\ Carlson, Y. C.\ Chen and M.\ Van-derhaeghen,
Phys.\ Rev.\ D {\bf 72}, 013008 (2005).

\bibitem{iwasaki}
Y.\ Iwasaki,
Prog.\ Theor.\ Phys.\ {\bf 46}, 1587 (1971).

\bibitem{sucher88}
G.\ Feinberg and J.\ Sucher,
Phys.\ Rev.\ D {\bf 38}, 3763 (1988).

\bibitem{holstein}
B. R.\ Holstein and J. F.\ Donoghue,
Phys.\ Rev.\ Lett.\ {\bf 93}, 201602 (2004)
[arXiv:hep-th/0405239];
B.R.\ Holstein and A.\ Ross,
arXiv hep-ph0802.0715. 

\bibitem{1}
Conventionally, one extracts a factor of $\sqrt {\frac{m}{E_p}} $ for each participating spin-1/2 particle and $\sqrt {\frac{1}{2 E_p}} $ for each spin-0 in $| i \rangle$ and $ | f \rangle$ where $E_{p}= \sqrt{p^2+m^2}$.

\bibitem{2}
We avoid using the term ``potential'' since it is not uniquely defined beyond first order perturbation theory as noted by Holstein~\cite{holstein}.

\bibitem{cohen} 
P.\ Avan, C.\ Cohen-Tannodji, J.\ Dupont, C.\ Fabre,
J.\ Phys.\ B {\bf 11}, 563 (1987).

%\bibitem{weinberg65}
%S.\ Weinberg, 
%Phys.\ Rev.\ {\bf140},  B516 (1965).

\bibitem{malta}
P.C.\ Malta, L.P.R.\ Ospedal, K.\ Veiga, and J.A.\ Helay$\ddot{e}$l-Neto,
Adv.\ High Energy Phys.\ 2531436 (2016). 

\bibitem{Fischback}
S.H.\ Aronson, Hai-Yang Cheng, Ephraim Fischback and Wick Haxton,
Phys.\ Rev.\ Lett.\ {\bf 56}, 1342 (1986).

\bibitem{Fayet}
P.\ Fayet, 
Phys.\ Lett.\ {\bf 95 B}, 285 (1980). 

\bibitem{Long03} \label{ref:Long03} 
J. C.\ Long, et al., Nature {\bf 421}, 922 (2003).
%; arXiv:hep-ph/0210004.

\bibitem{Yan14}
H.\ Yan, et al., Class.\ Quantum Grav. {\bf 31}, 205007
(2014).
% arXiv:1402.0145. 

\bibitem{Long15} \label{ref:Long15} 
J. C.\ Long and V. A.\ Kosteleck\'y,
Phys.\ Rev.\ D {\bf 91}, 092003 (2015).
% arXiv:1412.8362. 

\bibitem{Shao16}
C.-G.\ Shao, et al., Phys.\ Rev.\ Lett.\ 117 071102 (2016).
% arXiv:1607.06095. 
\end{thebibliography}

\end{document}